\documentclass[10pt,journal]{IEEEtran}
\IEEEoverridecommandlockouts
\usepackage[T1]{fontenc}
\usepackage{amssymb}
\usepackage{longtable}
\usepackage{amsmath}
\usepackage{amsthm}
\usepackage{amsfonts}
\usepackage{amsthm}
\usepackage{bbm}
\usepackage{amsmath}
\usepackage{amsfonts}
\usepackage{xcolor}
\usepackage{xcolor}
\usepackage{algorithmic}
\usepackage{algorithm}
\usepackage{array}
\usepackage{textcomp}
\usepackage{stfloats}
\usepackage{url}
\usepackage{verbatim}
\usepackage{graphicx}
\usepackage{cite}
\usepackage{balance}
\usepackage{float}
\usepackage{hyperref}
\usepackage{enumitem}
\usepackage{multirow}
\usepackage{booktabs}
\usepackage{xfrac}
\usepackage{amsmath}   
\usepackage{graphicx}  
\usepackage{caption}
\usepackage{array}
\usepackage{bm}
\usepackage{scalerel}

\newcolumntype{C}[1]{>{\centering\arraybackslash}p{#1}} 

\newtheorem{theorem}{Theorem}

\newtheorem{cor}[theorem]{Corollary}

\newtheorem*{blub}{Definition}
\newtheorem*{bla}{Remark}

\DeclareRobustCommand
  \Compactcdots{\mathinner{\cdotp}}

\begin{document}

\title{Maximum Spectral Efficiency With Adaptive MQAM Transmissions Over Terrestrial Coherent FSO Links}
\author{
\IEEEauthorblockN{Himani~Verma, Kamal~Singh, and Ranjan~K.~Mallik,~\IEEEmembership{Fellow,~IEEE}}
\vspace{-0.75cm}
\thanks{{Himani~Verma and Kamal~Singh are with the Department of Electrical Engineering, Shiv Nadar Institution of Eminence, Delhi NCR, 201314, India (e-mail: \tt{hv790@snu.edu.in}; \tt{kamal.singh@snu.edu.in}).}
{Ranjan~K.~Mallik is with the Department of Electrical Engineering, Indian Institute of Technology, Delhi, Haus Khas, New Delhi 110016, India (email: \tt{rkmallik@ee.iitd.ernet.in}).}}
}

\maketitle

\begin{abstract}
Coherent free-space optical (FSO) communication is recognized as a key enabler for ultra-high-capacity fronthaul and backhaul links in next-generation wireless networks. Spectrally efficient $M$--ary quadrature amplitude modulation (MQAM) formats are well-suited for these links. However, theoretical analyses of adaptive MQAM transmissions over terrestrial FSO channels remain limited. In this letter, we first derive the spectral efficiency limit of adaptive unconstrained MQAM \textcolor{black}{over gamma-gamma turbulence with pointing error}. We then show that adaptive transmissions using only six square MQAM constellations performs close to the theoretical limit (within $\bm{0.10\text{--}0.12}$ bits/s/Hz) across a wide range of signal-to-noise ratios and channel conditions.
\end{abstract}

\begin{IEEEkeywords}
Free-space optical (FSO) communication, adaptive transmission, average spectral efficiency, $M$-ary quadrature amplitude modulation (MQAM), gamma-gamma turbulence.
\end{IEEEkeywords}

\vspace{-0.1cm}
\section{Introduction}

Current commercial free-space optical (FSO) systems, like early fiber-optic implementations, predominantly employ intensity modulation with direct detection (IM-DD) due to their simple transceiver design. These IM-DD FSO systems typically support data rates in the range of $1$$-$$10$ Gbps per channel \cite{ecsystem2021}. Although coherent FSO systems have demonstrated superior performance in prior studies, their commercial adoption has been limited by increased complexity and cost \cite{barry90}. By contrast, novel advancements in digital signal processing (DSP) capabilities by the mid-2000s made coherent techniques commercially viable in fiber-optic systems, enabling their widespread deployment for high spectral efficiency and long-distance transmission without signal regeneration \cite{kikuchi2016}.

The explosive increase in demand for ultra-high-speed data transmission in 5G and emerging 6G networks, coupled with stringent fronthaul and backhaul requirements, has prompted renewed focus on coherent communication for terrestrial FSO links. This research field is still at a formative stage, with the majority of existing work being experimental and focused on the development of first-generation coherent FSO transceivers \cite{estol2023}. Recent terrestrial field trials have validated the feasibility of ultra-high-speed coherent FSO communication: i) at short to moderate distances ($< 1$ km), achieving data rates up to 100 Gbps per channel over $650$ m \cite{walsh2022}, $400$ Gbps per channel over $220$ m \cite{matsuda2021}, with peak rates exceeding $800$ Gbps per channel over $42$ m \cite{Guiomar2022}; and ii) at longer distances ($\geq 1$ km), achieving $120$ Gbps per channel over $1$ km \cite{feng2018}, $200$ Gbps per channel over $10.45$ km \cite{dochhan2019}, and most recent $400$ Gbps per channel coherent transmission over a $1.8$ km FSO link \cite{marco2024main}.

These high-throughput coherent FSO prototypes, similar to modern fiber-optic systems, employ $M$-ary quadrature amplitude modulation (MQAM) for its high spectral efficiency and straightforward I/Q implementation \cite{walsh2022,matsuda2021,Guiomar2022,feng2018,dochhan2019, marco2024main}. In particular, \cite{matsuda2021}, \cite{Guiomar2022} and \cite{marco2024main} implement adaptive probabilistic-constellation-shaped MQAM transmission, whereas \cite{walsh2022}, \cite{feng2018} and \cite{dochhan2019} rely on fixed MQAM formats. This growing preference for MQAM in coherent FSO research also parallels recent trends in fiber-optic communications \cite[Fig. 1]{alvarado2018}. Many of these field trials, especially those employing adaptive MQAM, demonstrate that advanced DSP developed for coherent fiber-optic systems can be effectively adapted to terrestrial FSO links. However, the distinctive impairments of terrestrial propagation continue to impose substantial DSP and implementation challenges for coherent FSO systems, as discussed by Guiomar \emph{et al.}~\cite{Guiomar2024dsp}.

\textcolor{black}{A critical research gap in the coherent FSO literature is
the lack of rigorous theoretical analysis on the fundamental
limits of communication achievable with adaptive MQAM transmissions, particularly under practical channel impairments such as 
\emph{strong turbulence} and \emph{pointing errors} in terrestrial links. We address this gap by developing a theoretical framework for a coherent FSO system that utilizes adaptive MQAM transmissions to enable \emph{reliable and spectrally efficient} operation, while explicitly accounting for these practical terrestrial impairments.} The analysis considers unconstrained MQAM constellations with uniformly distributed signals and derives the maximum spectral efficiency limit through joint adaptation of modulation order and transmit power under a fixed bit error rate (BER) constraint. \textcolor{black}{The main contributions are as follows:
\begin{itemize}[leftmargin=3mm,topsep=0pt,partopsep=0pt]
\item We derive an exact closed-form expression for the average spectral efficiency (ASE) limit of adaptive unconstrained MQAM transmissions over terrestrial coherent FSO links affected by gamma-gamma turbulence and pointing errors.
\item We then rigorously confirm the accuracy of the ASE limit through extensive Monte Carlo simulations over varying channel conditions and signal-to-noise ratios (SNRs).
\item Finally, we demonstrate that a practical finite set of square MQAM constellations achieves spectral efficiencies very close to the theoretical ASE limit over the $[0,30]$ dB SNR range under varied turbulence and pointing-error conditions.
\end{itemize}}

\noindent
\emph{Paper Organization:} Section \ref{sec:systemmodel} sets up the adaptive MQAM-based coherent FSO system and terrestrial channel models. Section \ref{sec:ASElimit} derives the ASE limit for adaptive unconstrained MQAM, motivating the adaptive transmission based on a compact set of square constellations in Section~\ref{sec:finiteMQAM}. Section \ref{sec:numericResults} demonstrates the scheme's near-optimal performance, and Section \ref{sec:conclusion} concludes with hints at future research directions.

\section{FSO System and Channel Models}\label{sec:systemmodel}
\emph{(i) Terrestrial FSO Link}: We consider a \textit{medium-haul} optical link supported by a good-precision acquisition, tracking, and pointing (ATP) subsystem, in which \textcolor{black}{irradiance fluctuations are mainly due to atmospheric turbulence and pointing errors~\cite{andrews2005laser}. The overall irradiance fluctuation $I$ is modeled as
\begin{align}\label{eq:overallChannelgain}
I \,=\, I_{\mathrm{a}} I_{\mathrm{p}},
\end{align}
where $I_{\mathrm{a}}$ and $I_{\mathrm{p}}$ are the atmospheric turbulence--induced and pointing-error--induced fluctuations, respectively.} The statistics of $I_{\mathrm{a}}$ follow the widely adopted gamma-gamma (GG) distribution for weak-to-strong turbulence conditions~\cite{andrews2005laser}, given by
\begin{equation}\label{eq:gg_dist}
f_{I_{\mathrm{a}}}(I_{\mathrm{a}}) =C \cdot  I_{\mathrm{a}}^{{{(\alpha + \beta)/2} -1}} K_{\alpha - \beta }\bigl( 2\sqrt{\alpha \beta I_{\mathrm{a}}}\bigr);\,\,\,I_a \geq 0,
\end{equation}
where $C := {(2(\alpha \beta)^{{(\alpha + \beta)}/{2}})}/({\Gamma ( \alpha) \Gamma ( \beta)})$, $\Gamma(\cdot)$ is the gamma function \cite[(8.310.1)]{gradshteyn2000table}, $K_{l} (\cdot)$ is the modified Bessel function of the second kind and order $l$ \cite[(8.494.1)]{gradshteyn2000table}, and the turbulence-dependent GG parameters $\alpha$ and $\beta$ are given by
\begin{align}
\alpha \,&=\, \bigl[ {\exp \bigl( {{0.49\sigma_{\mathrm{\scriptscriptstyle R}}^2}} {\bigl( {1 + 1.11\sigma_{\mathrm{\scriptscriptstyle R}}^{2.4}} \bigr)}^{-7/6} \bigr) - 1} \bigr]^{ - 1},\label{eq:GG_alpha}\\
   \beta \,&=\,  \bigl[ {\exp \bigl( {0.51\sigma_{\mathrm{\scriptscriptstyle R}}^2 \bigl( 1+0.69\sigma_{\mathrm{\scriptscriptstyle R}}^{2.4}\bigr)^{-5/6}}\bigr) - 1} \bigr]^{ - 1},\label{eq:GG_beta}
\end{align}
where, in turn, $\sigma_{\mathrm{\scriptscriptstyle R}}^2 \,:=\, 1.23\,C_{\mathrm{n}}^2 k_{\mathrm{w}}^{7/6}{L^{11/6}}$ is the Rytov variance, $k_{\mathrm{w}} \,=\, 2\pi /{\lambda_{\mathrm{w}}}$ is the optical wave number, ${\lambda_{\mathrm{w}}}$ is the wavelength, $L$ is the propagation distance, and $C_{\mathrm{n}}^2$ is the index-of-refraction structure parameter \cite{andrews2005laser}.

\textcolor{black}{In \cite{farid2007outage}, Farid and Hranilovic introduce a probability model for $I_\mathrm{p}$ under the independent and identically distributed (i.i.d.) zero-mean Gaussian assumptions for horizontal and vertical receiver sways (i.e., zero boresight and identical jitters), as
\begin{equation}\label{eq:jitterPDF}
f_{I_{\mathrm{p}}}( I_{\mathrm{p}}) \,=\, \dfrac{\xi ^{2}}{A_\mathrm{\scriptscriptstyle 0}^{\xi ^{2}}} I_{\mathrm{p}}^{\xi ^{2} -1}; \ \ \ 0\leqslant I_{\mathrm{p}} \leqslant A_\mathrm{\scriptscriptstyle 0},
\end{equation}
with distribution shaping parameters $A_\mathrm{\scriptscriptstyle 0}$ and $\xi$ defined therein. In this work, we adopt the recent \emph{modified intensity uniform model} proposed in~\cite{Miao2023newpointingmodel}, which preserves the distribution form of \eqref{eq:jitterPDF} but redefines the shaping parameters as
\begin{align}
A_\mathrm{\scriptscriptstyle 0} &\triangleq 1 - \exp(-2 r_\mathrm{\scriptscriptstyle A}^2 / w_\mathrm{\scriptscriptstyle L}^2),\label{eq:redefined_A0}\\
\xi^2 &\triangleq r_\mathrm{\scriptscriptstyle A}^2 /(2 \sigma_{\mathrm{e}}^2 A_\mathrm{\scriptscriptstyle 0}),\label{eq:redefined_xi}
\end{align}
where $r_\mathrm{\scriptscriptstyle A}$ is the aperture radius of receiver lens, $w_\mathrm{\scriptscriptstyle L}$ is the received beam waist, and $\sigma_{\mathrm{e}}^2$ is the pointing error variance. This refinement reduces the normalized mean-square error by over an order-of-magnitude relative to the Farid-Hranilovic model, yielding a more accurate pointing-error characterization.}

\textcolor{black}{Given the marginal distributions of the independent $I_{\mathrm{a}}$ and $I_{\mathrm{p}}$, the probability density function (PDF) of the overall $I$ is
\begin{align}\label{eq:PEPdf0}
f_{I}( I) \,&=\, \scaleto{\int}{3.95ex} f_{I|I_{\mathrm{a}}}( I|I_{\mathrm{a}}) f_{I_{\mathrm{a}}}( I_{\mathrm{a}}) dI_{\mathrm{a}}\,,	
\end{align}
where $f_{I|I_{\mathrm{a}}}\!\left(I|I_{\mathrm{a}}\right)$ denotes the conditional PDF, given by
\begin{equation}\label{eq:ConProb}
f_{I|I_{\mathrm{a}}}( I|I_{\mathrm{a}}) = \dfrac{\xi ^{2}}{A_\mathrm{\scriptscriptstyle 0}^{\xi ^{2}}I_{\mathrm{a}}} \left(\dfrac{I}{I_{\mathrm{a}}} \right)^{\xi ^{2} -1};  \ 0\leqslant I\leqslant A_\mathrm{\scriptscriptstyle 0} I_{\mathrm{a}}.
\end{equation}   
For $l \notin \mathbb{Z}$ and $|x| < \infty$, $K_{l}(x)$ has the series expansion \cite{gradshteyn2000table}:
\begin{align}\label{eq:besseKseries}
K_{l}(x) \!=\! \dfrac{\pi}{2 \! \sin(\pi l)} 
\sum_{k=0}^{\infty} \Biggl[
\frac{\left(\tfrac{x}{2}\right)^{2k - l}}{\Gamma(k \!-\! l \!+\! 1) k!} \!-\! \frac{\left(\tfrac{x}{2}\right)^{2k  +  l}}{\Gamma(k \!+\! l \!+\! 1) k!}\Biggr].
\end{align}
Inserting \eqref{eq:besseKseries} into \eqref{eq:gg_dist} and using the result with \eqref{eq:ConProb} in \eqref{eq:PEPdf0} yields
\begin{equation}\label{eq:PEPdf}
f_I(I)= \dfrac{\xi ^{2}}{A_\mathrm{\scriptscriptstyle 0}}\Biggl(\,\sum_{k=0}^\infty \, \sum_{x\in\{\alpha,\beta\}}\!S_k\bigl(x,\bar{x},\xi^2\bigr)
\left(\dfrac{I}{A_\mathrm{\scriptscriptstyle 0}}\right)^{k+\bar{x}-1}\Biggr),
\end{equation}
with the generalized power-series coefficients
\begin{align}\label{eq:sKcoeff}
S_k (x,\bar{x},\xi^2)\triangleq \dfrac{\mathrm{cosec}[\pi(\bar{x}\!-\!x)] \pi (x\bar{x})^{k+\bar{x}}}{ \Gamma(x)\Gamma(\bar{x})\Gamma(k\!-\!x\!+\!\bar{x}\!+\!1)(k\!+\!\bar{x}\!-\!\xi^2) k!},\!\!
\end{align} 
where $\bar{x}$ denotes the counterpart of $x$ (i.e., $\bar{\alpha} = \beta$ and $\bar{\beta} = \alpha$).}

\emph{(ii) Coherent FSO Transceiver}: Terrestrial FSO links are characterized by extremely large coherence bandwidths (hundreds of gigahertz) and millisecond-scale coherence times \cite[Ch.~18]{andrews2005laser}, and thus can be modeled as flat block-fading channels. This regime suits adaptive transmission, as the receiver (Rx) can accurately estimate the channel within each coherence block and feed back to the transmitter (Tx), similar to cellular mobile systems. For tractability, we assume perfect CSI at both Tx and Rx, fixed block durations, and i.i.d. GG fading across blocks. The coherent FSO Tx performs adaptive MQAM transmission, while the coherent Rx uses ideal heterodyne detection, yielding a complex baseband output as~\cite[Section~IV]{barry90}
\begin{equation}\label{eq:HD}
y \,=\,  h x_{\mathrm{\scriptscriptstyle QAM}} +w,
\end{equation}
where $h \!=\! |h|e^{-j\phi}  \! \in \! \mathbb{C}$ is the fading coefficient, $x_{\mathrm{\scriptscriptstyle QAM}}  \!\in \! \mathbb{C}$ the transmitted MQAM symbol, and  $w \!\sim \! \mathcal{CN}( 0,\sigma ^{2}_{\mathrm{\scriptscriptstyle OLO}})$ the (dominant) additive white Gaussian shot noise (AWGN) due to the optical local oscillator action. Since the synchronized receiver tracks $h$ perfectly, the channel-induced phase rotation $e^{-j\phi}$ is compensated and only $|h|$ is relevant for signal detection. As $h$ scales the transmitted optical field, the instantaneous channel intensity gain is given by $I \!:= \! | h| ^{2}$. Conditioned on $I$, the block-wise average electrical signal-to-noise ratio is given by
\begin{equation}\label{eq:SNR_HD}
 I P(I) /{\sigma _{\mathrm{\scriptscriptstyle OLO}}^{2}},
\end{equation}
where $P(I) := \mathbb{E}\left[| x_{\mathrm{\scriptscriptstyle QAM}}|^2\right]$ is the per-block average power for the chosen MQAM constellation. The long-term (over blocks) average transmit power is constrained as $\mathbb{E} [P(I)] = P_{\text{avg}}$.

\section{Maximum Spectral Efficiency with Adaptive Unconstrained MQAM Constellations}\label{sec:ASElimit}
Considered block-wise, the optical turbulence channel is a fixed-gain AWGN channel. A well-established bound on the probability of bit error (say, $P_b$) for MQAM transmission over an AWGN channel is presented in \cite[Section~9.3]{goldsmithbook2005}; we borrow this bound in our coherent FSO system model by accounting for the randomly time-varying block fading state $I$ as follows:
\begin{align}\label{eq:Pb_bound}
P_b (I) \,\leq \, 0.2 \, \mathrm{exp}\left(- \, \dfrac{1.5}{(M(I)-1)}\,\dfrac{I P(I)}{\sigma _{\mathrm{\scriptscriptstyle OLO}}^{2}}\right).
\end{align}
Here, $M(I)$ denotes the MQAM constellation size. Importantly, the bound in \eqref{eq:Pb_bound} is known to be `tight' (within $1$ dB) for $M \geq 4$ and received SNRs in the range of $[0,30]$ dB.

At the coherent FSO transmitter, the objective is to design an adaptive MQAM scheme that ensures \emph{reliable communication} in every fading block. This requires selecting the modulation order and transmit power on a per-block basis using the instantaneous channel state $I$, while guaranteeing a prescribed detection reliability. To enforce this, a target BER $P_B$ is imposed uniformly across all fading realizations, i.e., $P_b (I) = P_B,\,\forall I$. Enforcing this BER constraint and inverting \eqref{eq:Pb_bound} yields the adaptive MQAM constellation size as
\begin{align}\label{eq:mqam_const_size}
M(I)\,=\, 1 + \dfrac{1.5}{ - \ln (5P_B)} \dfrac{I P(I)}{\sigma _{\mathrm{\scriptscriptstyle OLO}}^{2}} \,=\, 1 + \mathcal{K} \dfrac{I P(I)}{\sigma _{\mathrm{\scriptscriptstyle OLO}}^{2}},
\end{align}
where $\mathcal{K} := -1.5/ \ln(5P_B)  < 1$. We assume that 
\begin{itemize}[leftmargin=3mm]
\item Each constellation $M(I)$ has $M(I)$ equiprobable symbols.
\item Each constellation employs ideal Nyquist pulse shaping.
\end{itemize}
With these assumptions and the fixed target BER, the adaptive MQAM achieves $\log_2 M(I)$ bits/s/Hz of spectral efficiency per block. The resulting average spectral efficiency (ASE) is
\begin{align}\label{eq:averageSE_with_MQAM}
\mathbb{E}\left[\log_2 M(I)\right] \,=\, \mathbb{E}\left[\log_2 \left(1 + \mathcal{K} {I P(I)}/{\sigma _{\mathrm{\scriptscriptstyle OLO}}^{2}} \right)\right].
\end{align}
Maximizing~\eqref{eq:averageSE_with_MQAM} over $P(I)$ subject to the power constraint $\mathbb{E}[P(I)] = P_{\text{avg}}$ is a simple convex program, whose solution yields the optimal MQAM power adaptation as
\begin{align}\label{eq:optimalPowAlloc}
\mathcal{K} P (I) / \sigma _{\mathrm{\scriptscriptstyle OLO}}^{2} \,=\, (I_{\text{th}}^{-1} - I^{-1})^{+},
\end{align}
where $a^{+} \triangleq \max\{a,0\}$, and $I_{\text{th}}$ is set by the power constraint. The $I_{\text{th}}$ enforces a channel cutoff below which transmission ceases: $P(I) = 0$ for $I < I_{\text{th}}$. Substituting~\eqref{eq:optimalPowAlloc} into~\eqref{eq:averageSE_with_MQAM} gives
\begin{align}\label{eq:123112}
&\mathbb{E}\left[\log_2 M(I)\right]\,=\, \mathbb{E}\left[(\ln\, ({I}{I^{-1}_{\text{th}}}))^{+}\right]/\,{\ln 2}.
\end{align}

\begin{blub}
The average SNR is defined as $\mathrm{SNR} \triangleq\! {P_{\mathrm{avg}}}/{\sigma _{\mathrm{\scriptscriptstyle OLO}}^{2}}\cdot$
\end{blub}
The ASE is frequently denoted by ${R}/{B}$ in the literature, and henceforth we will use this notation. Our main result follows.

\begin{theorem}\label{eq:thm1}
\textcolor{black}{The maximum ASE achievable by the adaptive unconstrained MQAM transmissions over the GG turbulence channel with pointing errors \textcolor{black}{described in \eqref{eq:overallChannelgain}}, is given by
\begin{align}\label{eq:ASEMQAM_ExactSeries}
\dfrac{R}{B} = \dfrac{1}{\ln 2}&\biggl[\,\ln\left(\frac{ A_\mathrm{\scriptscriptstyle 0}}{\alpha \beta\, I_{\mathrm{th}}}\right) + \,\psi (\alpha) \,+\, \psi (\beta) \,-\, \dfrac{1}{\xi^2} \notag \\
\,\,\,\,\phantom{xx}+ \xi^2 &\biggl(\,\sum_{k=0}^\infty  \, \sum_{x\in\{\alpha,\beta\}} \!\! \dfrac{S_k(x,\bar{x},\xi^2)}{ (k+\bar{x})^2}\! \left(\dfrac{I_{\mathrm{th}}}{A_\mathrm{\scriptscriptstyle 0}}\right)^{\!\!k + \bar{x}}\,\biggr)\,\biggr],
\end{align}
where $\alpha$, $\beta$, $A_\mathrm{\scriptscriptstyle 0}$, $\xi^2$, and $S_k(x,\bar{x},\xi^2)$ coefficients are defined in \eqref{eq:GG_alpha}, \eqref{eq:GG_beta}, \eqref{eq:redefined_A0}, \eqref{eq:redefined_xi}, and \eqref{eq:sKcoeff}, respectively, $\psi(\cdot)$ is the Digamma function, and $I_{\mathrm{th}}$ is solved from $\mathbb{E} [({I^{-1}_{\mathrm{th}}} - {I}^{-1})^{+}] \,=\, \mathcal{K} \,\!\,(\mathrm{SNR})$.}
\end{theorem}
\begin{IEEEproof}
\textcolor{black}{The expectation in \eqref{eq:123112} is simplified as
\begin{align}
&\mathbb{E}\left[\ln  ({I}{I^{-1}_{\text{th}}})^{+}\right] =\,\mathbb{E}\left[\ln  {I} \Compactcdots \mathbbm{1}_{I \geq I_{\text{th}}} \right] + \mathbb{E}\left[\ln  {I}^{-1}_{\text{th}} \Compactcdots \mathbbm{1}_{I \geq I_{\text{th}}} \right]\!,\\
&\phantom{xcdx}=\underbrace{\mathbb{E}\left[\ln  {I}\right]}_{= \, I_{1}} - \underbrace{\mathbb{E}\left[\,\ln  {I} \Compactcdots \mathbbm{1}_{I < I_{\text{th}}}\right]}_{= \, I_{2}} +\underbrace{\mathbb{E}\left[\ln  {I^{-1}_{\text{th}}} \Compactcdots \mathbbm{1}_{I \geq I_{\text{th}}}\right]}_{= \, I_{3}},\label{eq:I1I2I3}\notag\\[-2em]
\end{align}
where $\mathbbm{1}_{(\text{condition})}$ is the indicator ($1$ when the condition holds, $0$ else). Using the identity $(d/dn)\,x^{n}=x^{n}\ln x$, it follows that
\begin{equation}\label{eq:1stintegral33}
\mathbb{E} \left[ I^{n} \ln( I)\right]|_{n=0} \,=\, \mathbb{E}\left[(d/dn)\, I^{n} \right]|_{n=0}. 
\end{equation} 
The LHS in \eqref{eq:1stintegral33} is $\mathbb{E}\,[\ln (I)]$, i.e. $I_1$, while the RHS satisfy
\begin{equation}\label{eq:regularity}
\mathbb{E}\left[ (d/dn) \, I^n \right] = (d/dn)\, \mathbb{E}[I^n].
\end{equation} 
\eqref{eq:regularity} holds under standard regularity conditions. For the composite fading $I$, the $n$-th moment is given by (see \!\eqref{eq:gg_dist} and \!\eqref{eq:jitterPDF})
\begin{equation}\label{eq:moment}
\mathbb{E}[I^n] =\mathbb{E}[I_{\mathrm{a}}^n]\,\mathbb{E}[I_{\mathrm{p}}^n] =  
\frac{\Gamma\! \left( \alpha+ n\right) \Gamma\! \left( \beta + n\right)}{\alpha^n \beta^n \, \Gamma (\alpha)\,\Gamma (\beta)}\,  \frac{ A_\mathrm{\scriptscriptstyle 0}^n \, \xi^{2}}{n + \xi ^{2}}.
\end{equation}
Upon differentiation of \eqref{eq:moment} w.r.t. $n$ followed by evaluating the attained expression for $n = 0$, the final $I_{1}$ expression is
\begin{align}
I_1 \,=\, \ln\left(\frac{A_\mathrm{\scriptscriptstyle 0}}{\alpha \beta}\right) + \psi (\alpha) +\psi (\beta) - \frac{1}{\xi ^{2} } \,
\cdot
\end{align}
\noindent
The remaining part in \eqref{eq:I1I2I3} is simplified to
\begin{align}
I_3 \!-\! I_2 &= \int_{0}^{I_{\text{th}}} \dfrac{1}{y} \biggl(\int_{0}^{y} f_I(u)du\biggr)dy - \ln {I_{\text{th}}},\,\,\phantom{xcccdxxx}\label{eq:i3i2}\\
\intertext{which, upon direct evaluation after substituting \eqref{eq:PEPdf}, yields}
&= \xi^2 \! \Biggl[\sum_{k=0}^\infty  \sum_{x\in\{\alpha,\beta\}}  \!\!\!\dfrac{S_k(x,\bar{x},\xi^2)}{ (k+\bar{x})^2} \!\left(\!\dfrac{I_{\mathrm{th}}}{A_\mathrm{\scriptscriptstyle 0}}\!\right)^{k \!+\! \bar{x}}\Biggr] \!-\! \ln {I_{\text{th}}}.\label{eq:i3minusi2}
\end{align}
Substituting $I_1$ and $(I_3 \!- I_2)$ into \eqref{eq:I1I2I3} completes the proof.}
\end{IEEEproof}
\textcolor{black}{The ASE performance of adaptive MQAM over the pure GG turbulence channel (i.e., no pointing errors) is a special case of~\eqref{eq:ASEMQAM_ExactSeries}, and serves as a useful baseline for comparison.}
\begin{theorem}\label{eq:thm2}
\textcolor{black}{The maximum ASE achievable with the adaptive unconstrained MQAM transmissions over the GG turbulence channel `without' pointing errors described in \eqref{eq:gg_dist}, is given by
\begin{align}\label{eq:ExactASEwithoutPE}
\dfrac{R}{B} \,=\, \dfrac{1}{\ln 2}&\biggl[\,\ln\left(\frac{1}{\alpha \beta\, I_{\mathrm{th}}}\right) + \,\psi (\alpha) \,+\, \psi (\beta) \notag \\
\,\,\,\,\phantom{xx}+ \,\,&\biggl(\,\sum_{k=0}^\infty \,\, \sum_{x\in\{\alpha,\beta\}}  \dfrac{a_k(x,\bar{x})}{ (k+\bar{x})^2} \,\,I_{\mathrm{th}}^{k + \bar{x}}\,\biggr)\,\biggr],
\end{align}
where $a_k(x,\bar{x}) \triangleq \tfrac{\mathrm{cosec}[\pi(x - \bar{x})] \pi (x\bar{x})^{k+\bar{x}}}{ \Gamma(x)\Gamma(\bar{x})\Gamma(k-x+\bar{x}+1)k!}$, and $I_{\mathrm{th}}$ is solved from $\mathbb{E} [({I^{-1}_{\mathrm{th}}} - {I}^{-1})^{+}] \,=\, \mathcal{K} \,\!\,(\mathrm{SNR})$.}
\end{theorem}
\begin{IEEEproof} \textcolor{black}{The composite distribution \eqref{eq:PEPdf} converges to the GG distribution \eqref{eq:gg_dist} for the limiting values $A_\mathrm{\scriptscriptstyle 0} \!\! \to \! 1$ and $\xi^2 \!\! \to \! \infty$. Applying these limits to \eqref{eq:PEPdf} and \eqref{eq:ASEMQAM_ExactSeries} completes the proof.}
\end{IEEEproof}
\begin{cor}\label{eq:cor3}
\textcolor{black}{At high SNRs, $I_{\mathrm{th}} \!\approx\!  1/\mathcal{K}(\mathrm{SNR})$, \textcolor{black}{so the power series in \eqref{eq:ASEMQAM_ExactSeries} vanishes and the maximum ASE limit reduces to}
\begin{align}\label{eq:ASE_ExactwithPE2}
\dfrac{R}{B} \approx \dfrac{1}{\ln 2} \left[\ln\left(\frac{\mathcal{K} A_\mathrm{\scriptscriptstyle 0}}{\alpha \beta }\right) \!+ \!\psi (\alpha)\! + \!  \psi (\beta) \!- \!\dfrac{1}{\xi^2} \!+ \!\ln \mathrm{SNR}\right].
\end{align}
Thus, the ASE increases logarithmically with SNR, with an offset determined by the BER target and the GG and pointing-error distribution parameters. For the pure GG turbulence channel without pointing errors, \eqref{eq:ASE_ExactwithPE2} follows by applying the limits $A_\mathrm{\scriptscriptstyle 0} \to 1$ and $\xi^2 \to \infty$.}
\end{cor}
\begin{cor}\label{eq:cor4}
\textcolor{black}{The high-SNR ASE penalty due to pointing errors for adaptive unconstrained MQAM transmissions over the GG turbulence is
\begin{equation}\label{eq:HighSNRLoss}
\frac{1}{\ln 2}\biggl[\,\frac{1}{\xi ^{2}} \,-\, \ln A_\mathrm{\scriptscriptstyle 0}\,\biggr].
\end{equation}
This follows from Corollary~\ref{eq:cor3} by comparing the high-SNR ASEs under GG turbulence with and without pointing errors.}
\end{cor}
\begin{bla}\label{eq:remark1}
Interestingly, if we imagine setting $\mathcal{K}=1$, the expectation in \eqref{eq:averageSE_with_MQAM} coincides with the Shannon ASE limit of the terrestrial coherent FSO channels \cite{verma2024new}. Hence, the actual $\mathcal{K}$ can be interpreted as the SNR penalty due to uncoded uniformly-distributed MQAM signaling, when compared to perfectly adaptive and optimally coded transmissions. Alternatively, $\mathcal{K}$ can be viewed as an upper bound on the achievable coding gain with adaptive MQAM transmission.
\end{bla}

\section{Average Spectral Efficiency With Finite Square MQAM Constellations Set}\label{sec:finiteMQAM}
Section~\ref{sec:ASElimit} presented the ideal adaptive MQAM transmission scheme with continuously scalable $M$, whose practical realization is difficult~\cite{forney1984}. Following the discrete-rate approach of~\cite{goldsmithbook2005}, we restrict $M$ to a finite ordered set. As we will see in Section~\ref{sec:numericResults}, the resulting adaptive square MQAM transmission scheme---outlined next---attains rates close to the ASE limit.

For the unrestricted optimum, \eqref{eq:123112} gives $M(I) = (I/I_{\text{th}})^{+}$, i.e., $M(I)\!\propto\! I$ for $I \!> \! I_{\text{th}}$. In the discrete case, let $\{M_i\}_{i=0}^{N-1}$ be the admissible constellation sizes (ascending), with the $I$-range partitioned accordingly. We use square constellations $M_i\!=\!2^{2i}$ for $i \ge1$, and $M_0=0$ for $I\le I_{\text{th}}^{*}$. The transmit power in partition $i\geq1$ satisfying the target BER follows from \eqref{eq:mqam_const_size} as
\begin{align}
{P_i(I)} \,=\,  {(M_i  - 1)}{\sigma _{\mathrm{\scriptscriptstyle OLO}}^{2}}/{{\mathcal{K}} I},
\end{align}
with $P_0(I)=0$. A six-constellation example appears in Table~\ref{tab:finiteconstl}.
\begin{table}[h!]
\scriptsize
\caption{\small Rate and Power adaptation}
\centering
\begin{tabular}{c | c | c | c}
\hline\\[-1.025em] 
Partition $i$  &  $I$ Range  &  $M_i$ & ${P_i (I)}/{ \sigma _{\mathrm{\scriptscriptstyle OLO}}^{2}}$ \\[0.1em]
\hline
 &  & & \\[-1.05em]
$0$ & $0 \leq I/I_{\text{th}}^{*} < 4$ & $0$ & $0$\\[0.1em]\hline
 &  & & \\[-1.05em]
$1$ & $4 \leq I/I_{\text{th}}^{*} < 16$ & $4$ & ${3}/{{\mathcal{K}} I}$\\[0.1em]\hline
 &  & & \\[-1.05em]
$2$ & $16 \leq I/I_{\text{th}}^{*} < 64$ & $16$ & ${15}/{{\mathcal{K}} I}$\\[0.1em]\hline
 &  & & \\[-1.05em]
$3$ & $64 \leq I/I_{\text{th}}^{*} < 256$ & $64$ & ${63}/{{\mathcal{K}} I}$\\[0.1em]\hline
 &  & & \\[-1.05em]
$4$ & $256 \leq I/I_{\text{th}}^{*} < 1024$ & $256$ & ${255}/{{\mathcal{K}} I}$\\[0.1em]\hline
 &  & & \\[-1.05em]
$5$ & $1024 \leq I/I_{\text{th}}^{*} < \infty$ & $1024$ & ${1023}/{{\mathcal{K}} I}$\\[0.1em]\hline
\end{tabular}
\label{tab:finiteconstl}
\end{table}
Since the continuous-rate optimum satisfies $M(I)I_{\text{th}}=I$, the discrete boundaries are chosen as $M_iI_{\text{th}}^{*}$, giving
\begin{align}\label{eq:discrateQAM_orders}
M_iI_{\text{th}}^{*}\le I < M_{i+1}I_{\text{th}}^{*};\,\,\,i=0,\dots,N-1,
\end{align}
with $M_N:=\infty$. The achievable spectral efficiency is
\begin{align}\label{eq:discrateQAM_rate}
{R}/{B} = \textstyle{\sum_{i=1}^{N-1}} \{\log_2 M_i \} \, \mathrm{Pr}\! \left[ M_i \leq {I}/{I_{\text{th}}^{*}} < M_{i+1} \right].
\end{align}
The parameter $I_{\text{th}}^{*}$ \textcolor{black}{must satisfy the long-term power constraint:}
\begin{align}\label{eq:discrateQAM_powercons}
\textstyle{\sum_{i=1}^{N-1} \int_{M_i \textcolor{black}{I_{\text{th}}^{*}}}^{M_{i+1}  \textcolor{black}{I_{\text{th}}^{*}}}}\,\,(M_i - 1) I^{-1}  f_I (I) dI = \mathcal{K} (\mathrm{SNR}).
\end{align}

\begin{table*}[b]
\begin{center}
\caption{\small \textcolor{black}{Required Transmit SNR to Achieve the Specified ASE  Under a $10^{-3}$ BER Reliability Requirement.}}\label{tab:ReqSNR}
\scriptsize
\textcolor{black}{
\begin{tabular}{|c|c|c|c|c|c|c|c|c|}
\hline
\multirow{4}{*}{\begin{tabular}[c]{@{}c@{}}$\,\,\dfrac{R}{B}\,\,$\\[0.7em]\,\,(bits/s/Hz)\,\,\end{tabular}} 
& \multicolumn{8}{c|}{\rule{0pt}{2.15ex} Required normalized Transmit power: $\mathrm{SNR} = {P_{\mathrm{avg}}}/{\sigma _{\mathrm{\scriptscriptstyle OLO}}^{2}}$ (dB scale)} \\ 
\cline{2-9}
& \multicolumn{4}{c|}{GG atmospheric turbulence only} 
& \multicolumn{4}{c|}{GG atmospheric turbulence with Pointing errors} \\   
\cline{2-9}
& \multicolumn{2}{c|}{\rule{0pt}{2.25ex}\,\,\,Weak Turbulence: $\sigma_{\textrm{R}}^2=0.4$\,\,\,}
& \multicolumn{2}{c|}{\rule{0pt}{2.25ex}\,\,\,Strong Turbulence: $\sigma_{\textrm{R}}^2=2$\,\,\,}
& \multicolumn{2}{c|}{\rule{0pt}{2.25ex}\,\,\,Weak Turbulence: $\sigma_{\textrm{R}}^2=0.4$\,\,\,}
& \multicolumn{2}{c|}{\rule{0pt}{2.25ex}\,\,\,Strong Turbulence: $\sigma_{\textrm{R}}^2=2$\,\,\,} \\
\cline{2-9}
& \,\,fixed MQAM\,\, & adaptive MQAM & \,\,fixed MQAM\,\, & adaptive MQAM 
& \,\,fixed MQAM\,\, & adaptive MQAM & \,\,fixed MQAM\,\, & adaptive MQAM \\[-0.06em] \hline
2  & 14.0 & 10.7 & 20.3 & 11.2 & 17.6 & 13.4 & 26.3 & 17.0 \\[-0.175em] \hline
4  & 21.0 & 17.9 & 27.3 & 18.9 & 24.6 & 20.7 & 33.2 & 24.8 \\[-0.175em] \hline
6  & 27.2 & 24.2 & 33.5 & 25.4 & 30.9 & 27.0 & 39.5 & 31.2 \\[-0.175em] \hline
8  & 33.3 & 30.3 & 39.6 & 31.5 & 36.9 & 33.1 & 45.5 & 37.3 \\[-0.175em] \hline
10 & 39.3 & 36.3 & 45.6 & 37.5 & 43.0 & 39.1 & 51.6 & 43.4 \\[-0.06em] \hline
\end{tabular}}
\end{center}
\vspace{-0.77\baselineskip}
\end{table*}

\vspace{-0.45em}
\section{Numerical Results and Discussion}\label{sec:numericResults}
\textcolor{black}{For the numerical results, we consider a horizontal \emph{medium-range} terrestrial FSO link with fixed wavelength and propagation distance to isolate turbulence and jitter effects; key parameters are listed in Table~\ref{tab:FSO_System_settings}. The Rx beam waist $w_{\mathrm{\scriptscriptstyle L}}$ is estimated from the Tx beam waist $w_{\mathrm{\scriptscriptstyle 0}}$ via $\displaystyle w_{\mathrm{\scriptscriptstyle L}} \!\approx w_{\mathrm{\scriptscriptstyle 0}}( 1\!+\!\epsilon ( \lambda_{\mathrm{w}} L/\pi w_{\mathrm{\scriptscriptstyle 0}}^{2})^{2})^{1/2}$ with $\epsilon \!=\! ( 1+2w_{\mathrm{\scriptscriptstyle 0}}^{2} /\rho_{\mathrm{\scriptscriptstyle 0}}^{2})$ and $\rho_{\mathrm{\scriptscriptstyle 0}} \!=\! (1.46C_{\mathrm{n}}^{2} k_{\mathrm{w}}^{2} L)^{-3/5}$; see \cite[subsection~6.4.3]{andrews2005laser} for details. With $\lambda_{\mathrm{w}}$ and $L$  fixed, the turbulence strength---quantified by the Rytov variance $\sigma_\mathrm{\scriptscriptstyle R}^2$---is varied via $C_{\mathrm{n}}^2$ to span weak-to-strong turbulence conditions \cite[Section 12.2]{andrews2005laser}. The chosen parameters are representative of a medium-range terrestrial FSO link equipped with a well-designed ATP subsystem with `small' residual jitter.
\begin{table}[H]
\caption{\textcolor{black}{FSO System and Channel Settings}}\label{tab:FSO_System_settings}
\footnotesize
\centering
\textcolor{black}{
\begin{tabular}{@{} l   c  c @{}}
\toprule\\[-1.4em]
\textbf{Parameter}  &\textbf{Symbol}\,\,\,\,\,\,\, &\,\,\,\,\,\,\,\,\,\,\,\,\,\,\,\,\textbf{Value/Range}\\[-1.35em] \\ \midrule\\[-1.45em]
FSO link length&$L$\,\,\,\,\,\,\,&\,\,\,\,\,\,\,\,\,\,\,\,\,\,\,\,\,\,\,\,\,$\sfrac{1}{3}$ km\,\,\,\,\,\,\,\\[0.0ex]
Optical wavelength&$\lambda_{\mathrm{w}}$\,\,\,\,\,\,\,&\,\,\,\,\,\,\,\,\,\,\,\,\,\,\,\,\,\,\,\,\, 1550 nm\,\,\,\,\,\,\,\\[-0.25ex]
Tx beam waist&$w_\mathrm{\scriptscriptstyle 0}$\,\,\,\,\,\,\,&\,\,\,\,\,\,\,\,\,\,\,\,\,\,\,\,\,\,\,\,\, 1.5 cm\,\,\,\,\,\,\,\\[-0.25ex]
Rx aperture diameter&$2r_\mathrm{\scriptscriptstyle A}$\,\,\,\,\,\,\,&\,\,\,\,\,\,\,\,\,\,\,\,\,\,\,\,\,\,\,\,\, 4.0 cm\,\,\,\,\,\,\,\\[-0.2ex]
Refractive-index str. param.&$C_{\mathrm{n}}^2$\,\,\,\,\,\,\,&\,\,\,\,\,\,\,\,\,\,\,\,\,\,\,\,\,\,\,\,$10^{\textrm{-}13}$$-$$10^{\textrm{-}15}\,\,\text{m}^{\textrm{-}2/3}$\,\,\\[-0.2em]
\bottomrule\\[-1.475em]
\end{tabular}
\\[0.5em]\text{\centering{\textbf{\,\,\,\,\,\,\,\,GG turbulence without pointing error\phantom{x}\,\,\,\,\,\,\,\,\,\,}}}\vspace{0.1em}
\centering
\begin{tabular}{@{} l c c @{}}	
\toprule\\[-1.45em]
\textit{Turbulence\,\,\,\,\,\,\,\,\,\,\,\,\,}&{\,\,\,\,Rytov variance\,\,}&\textit{\,\,\,\,\,\,\,\,\,\,\,\,\,\,\,\,GG distribution\,\,\,\,\,\,}\\[-0.24em]
\textit{\,\,strength\,\,\,\,\,\,\,\,\,\,\,\,} &\,\,\,\,\,\,\,\,\,\,$\sigma_{\mathrm{\scriptscriptstyle R}}^2$\,\,\,\,\,\,\,\,\,\,\,\,\,\,&\textit{\,\,\,\,\,\,\,\,\,\,parameters} {(\eqref{eq:GG_alpha} and~\eqref{eq:GG_beta})}\\[-0.15em] \midrule\\[-1.45em]
\,Weak\,\,\,\,\,\,\,&\,\,\,\,\,\,\,\,\,$0.4$\,\,\,\,\,\,\,\,\,\,\,\,&\,\,\,\,\,\,\,\,\,\,\,$\alpha \,=\, 6.8755,\,\beta \,=\, 5.3384$\,\,\\[-0.25ex]
\,Moderate\,\,\,\,\,\,\,\,&\,\,\,\,\,\,\,\,\,$1$\,\,\,\,\,\,\,\,\,\,\,\,&\,\,\,\,\,\,\,\,\,\,\,$\alpha \,=\, 4.3939,\,\beta \,=\, 2.5636$\,\,\\[-0.25ex]
\,Strong\,\,\,\,\,\,\,\,&\,\,\,\,\,\,\,\,\,$2$\,\,\,\,\,\,\,\,\,\,\,\,&\,\,\,\,\,\,\,\,\,\,\,$\alpha \,=\, 3.9929,\,\beta \,=\, 1.7018$\,\,\\[-0.65ex]
\bottomrule\\[-1.1em]
\end{tabular}
\\[0.15em]\text{\centering{\textbf{\,\,\,\,\,\,GG turbulence with pointing error ($\sigma_{\mathrm{e}} = 1$ cm)\phantom{x}\,\,\,\,\,\,\,\,\,\,}}}\vspace{0.1em}
\centering
\begin{tabular}{@{} l c c @{}}
\toprule\\[-1.45em]
\textit{Turbulence\,\,\,\,\,\,\,\,\,\,\,\,\,}&{\,\,\,\,Rytov variance\,\,}&\textit{\,\,\,\,\,\,\,\,\,\,Pointing Error}\\[-0.2em]
\textit{\,\,strength\,\,\,\,\,\,\,\,\,\,\,\,} &\,\,\,\,\,\,\,\,\,\,$\sigma_{\mathrm{\scriptscriptstyle R}}^2$\,\,\,\,\,\,\,\,\,\,\,\,\,\,&\,\,\,\,\,\,\,\,\,\,\textit{dist. parameters} {(\eqref{eq:redefined_A0} and \eqref{eq:redefined_xi})}\\[-0.15em] \midrule\\[-1.45em]	
\,Weak\,\,\,\,\,&\,\,\,\,\,\,\,\,\,$0.4$\,\,\,\,\,\,\,\,\,\,\,\,&\,\,\,\,\,\,\,\,\,\,\,$\xi \,=\, 1.7808
,\,A_\mathrm{\scriptscriptstyle 0} = 0.7180
$\,\,\\[-0.25ex]
\,Moderate\,\,\,\,\,\,&\,\,\,\,\,\,\,\,\,$1$\,\,\,\,\,\,\,\,\,\,\,\,&\,\,\,\,\,\,\,\,\,\,\,$\xi \,=\, 2.0491
,\,A_\mathrm{\scriptscriptstyle 0} = 0.4948
$\,\,\\[-0.25ex]
\,Strong\,\,\,\,\,\,&\,\,\,\,\,\,\,\,\,$2$\,\,\,\,\,\,\,\,\,\,\,\,&\,\,\,\,\,\,\,\,\,\,\,$\xi \,=\, 2.5848
,\,A_\mathrm{\scriptscriptstyle 0} = 0.3025
$\,\,\\[-0.65ex]
\bottomrule\\[-1.0em]
\end{tabular}}
\vspace{-0.85em}
\end{table}
Fig.~\ref{fig:ExWithoutPE3} validates the exact ASE expressions in Theorems~\ref{eq:thm1} and~\ref{eq:thm2}, showing close agreement with Monte Carlo simulations across the $[0,30]$ dB SNR range and all turbulence conditions; the series solutions use the first $20$ terms. Consistent with Corollary~\ref{eq:cor4}, the high-SNR ASE penalty \emph{caused by pointing errors}---shown as downward-pointing blue arrowed lines---follows the predicted behavior: stronger turbulence increases beam-spread loss (reduced $A_{\scriptscriptstyle 0}$) but offers only marginal jitter relief (slight increase in $\xi$), leading to a net worsening of ASE.
\begin{figure}[H]
\centering
\includegraphics[scale = 0.7]{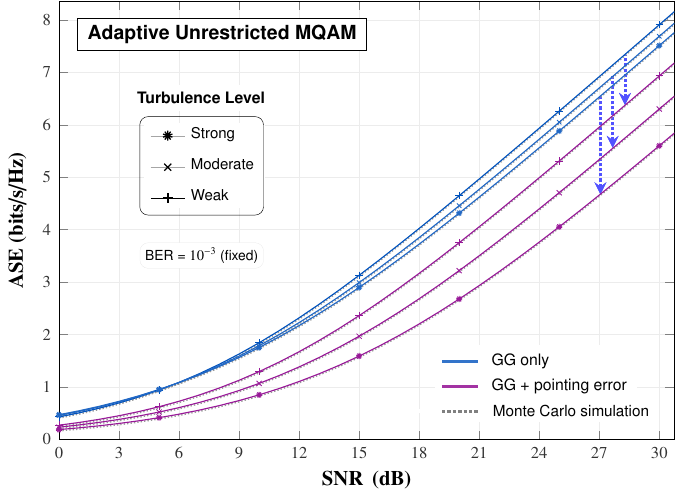}
\caption{\textcolor{black}{Numerical verification of the ASE limits for adaptive MQAM transmissions in Theorems~\ref{eq:thm1} and~\ref{eq:thm2} via Monte Carlo simulations ($4 \times 10^{4}$ realizations).}}
\label{fig:ExWithoutPE3}
\vspace{-1em}
\end{figure}}
\textcolor{black}{Table~\ref{tab:ReqSNR} summarizes the required SNRs for fixed and adaptive MQAM over GG turbulence channels, with and without pointing errors, to achieve specified ASE targets (or ${R}/{B}$) at $10^{-3}$ BER (or $P_B$). For a target ${R}/{B}$ at reliability $P_B$, the required SNR for adaptive MQAM is obtained numerically from \eqref{eq:ASEMQAM_ExactSeries}; for fixed MQAM, it is obtained numerically after taking expectations on both sides of \eqref{eq:Pb_bound} with $M(I) = {R}/{B}$, $P(I) = P_{\mathrm{avg}}$, and $\mathbb{E}[P_b(I)] = P_B$. The numerical results show that adaptive MQAM offers remarkable Tx power savings over fixed MQAM on GG channel with pointing errors (resp. without pointing errors): $8.2\text{--}9.3$~dB (resp. $8\text{--}9$~dB) under strong turbulence, and about $4$~dB (resp. about $3$~dB) under weak turbulence across the target ASE range. For a fixed Tx power budget, these savings roughly translate into $1\text{--}3$~bits/s/Hz ASE gains at high SNR, as per Corollary~\ref{eq:cor3}.}

\begin{figure}[H]
\centering
\includegraphics[scale = 0.67]{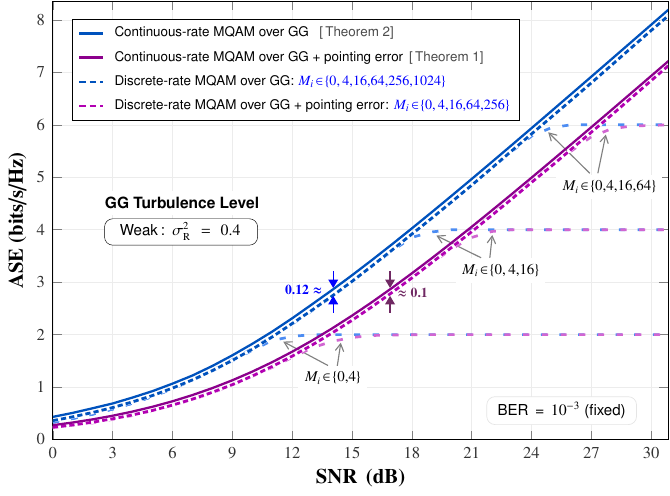}
\caption{\textcolor{black}{Average spectral efficiency of adaptive QAM transmissions over weak gamma-gamma turbulence channel with and without pointing errors.}}
\label{fig:ExWithoutPE4}
\vspace{-0.85em}
\end{figure}

\begin{figure}[H]
\centering
\includegraphics[scale = 0.67]{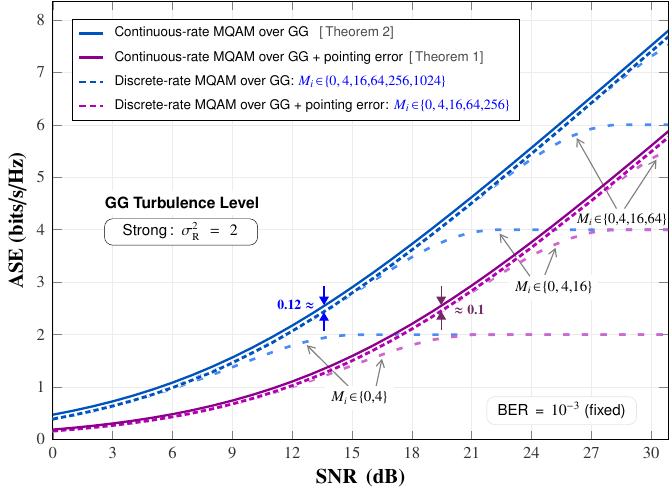}
\caption{\textcolor{black}{Average spectral efficiency of adaptive QAM transmissions over strong gamma-gamma turbulence channel with and without pointing errors.}}
\label{fig:ExWithoutPE6}
\vspace{-1em}
\end{figure}

\textcolor{black}{More importantly, we assess how closely the finite square-MQAM constellations based adaptive transmission scheme of Section~\ref{sec:finiteMQAM} can realize the theoretical ASE limit by evaluating \eqref{eq:discrateQAM_rate}. As shown in Figures~\ref{fig:ExWithoutPE4} and~\ref{fig:ExWithoutPE6}, the adaptive scheme—using only six constellations for GG channels without pointing errors and only five for GG channels with pointing errors—achieves spectral efficiencies close to the ASE limit (within $0.10\text{--}0.12$ bits/s/Hz) across a wide $[0,\!30]$ SNR range and channel profiles.}

\textcolor{black}{Similar power savings and near-optimal ASEs were also observed across target BERs ranging from $10^{-6}$ to $10^{-4}$.}

\balance
\textcolor{black}{These results provide concrete design guidelines for terrestrial coherent FSO links as follows.
\begin{itemize}[leftmargin=3mm,topsep=0pt,partopsep=0pt]
\item First, Table~\ref{tab:ReqSNR} quantifies the power margins required to compensate for turbulence and pointing errors. Under turbulence alone, strong turbulence imposes over $6$~dB additional SNR compared to weak turbulence for fixed MQAM but only $0.5\text{--}1.2$~dB for adaptive MQAM, underscoring its superior robustness. With pointing errors, the additional SNR margins increase for both fixed and adaptive MQAM (by comparable amounts), primarily to compensate for beam-spread loss.
\item Second, for a given Tx power budget, adaptive MQAM can deliver substantial ASE gains of up to $3$~bits/s/Hz over fixed MQAM at high SNRs in turbulence-dominated regimes.
\item Finally, the ASE results indicate that adaptive FSO links can achieve near-optimal spectrally efficient communication using low-complexity standard $4/16/64/256/1024$-QAM.
\end{itemize}}

\vspace{-0.5em}
\section{Conclusion}\label{sec:conclusion}
\textcolor{black}{In this paper, we have presented formulae for the ASE limit of adaptive MQAM-based coherent FSO communication over gamma-gamma turbulence channels, with and without pointing errors.} Although the theoretical limit assumes unrestricted MQAM, extensive numerical results show that it can be closely approached by an adaptive scheme employing a compact set of square MQAM constellations. This underscores the practical ease of implementing efficient adaptive MQAM and affirms the relevance of the derived ASE limit for terrestrial coherent FSO systems. The remaining gap to Shannon's ASE limit can be reduced via constellation shaping and advanced coding.

\bibliographystyle{IEEEtran}
\bibliography{references}
\end{document}